\documentclass[seceq]{ptptex}

\usepackage{graphicx}

\def\no{\nonumber \\}
\def\btab{\begin{table}[h] \begin{center} \begin{tabular}{l lp{3in}}}
      \def\etab{\end{tabular} \end{center} \end{table}}
\def\btabm{\begin{center} \begin{tabular}}
    \def\etabm{\end{tabular} \end{center}}

\def\a{{\alpha}}

\def\b{{\beta}}
\def\g{{\gamma}}
\def\G{{\Gamma}}
\def\d{\delta}

\def\ep{{\epsilon}}

\def\m{{\mu}}
\def\n{{\nu}}
\def\rh{\rho}

\def\t{{\theta}}

\def\om{\omega}

\def\f#1#2{{\frac{#1}{#2}}}

\def\s{\sqrt}

\def\f {\frac}

\def\p{\partial}

\def\CN{{\cal N}}
\def\CO{{\cal O}}

\notypesetlogo                       
\preprintnumber[3cm]{
KUNS-2184\\PTPTeX ver.0.9\\ January, 2008}

\title{
Warped AdS{\Large$_5$} Black Holes and Dual CFTs
}

\author{
Keiju \textsc{Murata,}\footnote{E-mail: murata@tap.scphys.kyoto-u.ac.jp}
Tatsuma \textsc{Nishioka}\footnote{E-mail: nishioka@gauge.scphys.kyoto-u.ac.jp}
and
Norihiro \textsc{Tanahashi}\footnote{E-mail: tanahashi@tap.scphys.kyoto-u.ac.jp}
}

\inst{%
Department of Physics, Kyoto University, Kyoto 606-8502, Japan
}

\abst{%
We consider a black hole solution whose spatial boundary is a squashed three sphere 
in Einstein gravity with negative cosmological constant.
We solve the Einstein equations numerically and find a warped AdS black hole solution with 
arbitrary squashing parameter.
This solution becomes the ordinary AdS-Schwarzschild solution when the squashing parameter is 
chosen appropriately. Motivated by this fact, we study $\CN =4$ super 
Yang-Mills theory with zero coupling constant on a squashed three sphere and show that the thermodynamical
entropy of this theory roughly agrees with that of the warped AdS black hole up to a factor of 3/4. 
We also study the confinement/deconfinement transition of the gauge theory. 
We evaluate the approximate Hawking-Page transition temperature of the warped AdS black hole
and find a qualitative agreement between the transition temperatures of the gravity and the gauge theory.
These results suggest a duality between the warped AdS solution and 
$\CN =4$ super Yang-Mills theory on a squashed three sphere. 
}

\PTPindex{121, 451}  

\begin{document}

\maketitle

\section{Introduction}
New black hole solutions in AdS$_5$ spacetime have been extensively studied in general relativity, and
they have played an important role in the recent developments of the AdS/CFT correspondence.\cite{Ma,GKP,Wi} \ 
One remarkable progress is the discovery of the confinement/deconfinement
transition in $\CN = 4$ super Yang-Mills theory (SYM)\cite{Su,AMMPV}, which was conjectured
to correspond to the Hawking-Page transition\cite{HP} in gravity theory by Witten.\cite{Wi2} \ 
Such transition and correspondence were well studied in various contexts%
\cite{HR,Liu,AMMPV2,AGLW,BaWa,KNY,YY,Y,HaOr,HaOr2,Hi,HiIi,HKO,
ABW,DG,HKNW,MNTY,FNT,HKO2,IIST,IKNT,KiMa}, and these results are regarded as 
evidence of the AdS/CFT correspondence.
It is also known that the thermodynamical quantities in free gauge theory agree
with those in dual gravity up to a factor of 3/4.\cite{Wi2,GKPe}

In this paper, we present a new duality between the black hole, which is asymptotically 
a squashed three sphere at spatial infinity, and the gauge theory defined
on its boundary similarly to the AdS/CFT correspondence.
We construct a warped black hole solution on AdS$_5$ and compare it with its dual gauge theory.

Such a deformation of the AdS/CFT correspondence is also suggested from the 
studies on the topological massive gravity (TMG) in three dimensions, 
which is defined by the Einstein-Hilbert action with the gravitational Chern-Simons 
term.\cite{DJT,DJT2} \ 
Black hole solutions called  warped 
AdS$_3$ black holes were constructed within TMG in Refs.~\citen{Nutku,Gurses,Clement}
and recently generalized in Ref.~\citen{ALPSS}.
It was conjectured that the warped AdS$_3$ black hole is dual to a two-dimensional 
CFT present on its boundary (see also Refs.~\citen{MCGL,CoDe,CDR,Ann,HHKNT} 
for related works).
These studies indicate the possibility of the deformation of the AdS/CFT correspondence
even in higher dimension.

By assuming a metric ansatz for a warped AdS$_5$ solution, the Einstein equations 
reduce to a set of ordinary differential equations. 
We solve them numerically because those equations are too complicated to solve analytically.
As a result, we obtain a two-parameter family of a warped black hole solution 
and a one-parameter family of a warped AdS solution.\footnote{
In four dimensions, such a distorted AdS black hole is constructed using
a static perturbation of the AdS-Schwarzschild black hole.\cite{Yos,Tomi}
}
Similar warped solutions were constructed for a zero cosmological constant 
case
in the quest for black hole solutions in compactified extra 
dimension.\cite{DM,GP,Sor,Rasheed,Larsen,IM}
$^,$\footnote{
Although it is difficult to introduce the cosmological constant to warped solutions in general, there are some partial results.
One example is the case of 
extreme squashed Kaluza-Klein black holes,\cite{IM}
to which
the positive cosmological constant was successfully introduced in Ref.~\citen{IIKMMT}.
}
The solutions we construct in this paper can be regarded as an extension
of those warped solutions with zero cosmological constant to 
that with negative cosmological constant.

The dual gauge theory should be $\CN=4$ SYM on a squashed $S^3$ because 
the warped solution becomes the ordinary AdS space when the squashing parameter is set appropriately.
We note that the matter contents of the gauge theory are the same as $\CN=4$ SYM, but 
there is no supersymmetry on the squashed $S^3$.
When we consider $\CN=4$ SYM on $R\times M_3$, where $M_3$ is an arbitrary three-dimensional manifold, 
there is no supersymmetry in the theory unless $M_3$ is a maximally symmetric space. The proof is 
given in the Appendix.

The organization of this paper is as follows.
We provide the metric ansatz and the Einstein equations for the warped
AdS solution in \S\ref{Sec:Ansatz}.
We construct
the numerical solution in \S\ref{Sec:Numerical}.
Even though we do not know the analytical form, we can calculate the temperature and entropy
of the warped AdS black holes.
In \S\ref{Sec:Gauge}, we study $\CN =4$ SYM on a squashed $S^3$, which is expected to 
be dual to the warped AdS solution. 
We summarize the spectrum in this theory and then calculate the effective action.
We can compute the entropy using this action and compare it with that of the warped AdS solution.
We will find that the ratio between them is nearly 3/4 for any squashing parameter.
This result supports our expectation that $\CN =4$ SYM on a squashed $S^3$
is dual to the warped AdS solution.
\S\ref{Sec:discussion} is devoted to the discussion.

\section{Metric Ansatz and the Einstein Equations}
\label{Sec:Ansatz}
We parametrize the metric as
\begin{equation}
  ds^2= - F(r)e^{-2\delta (r)} dt^2
+\frac{d r^2}{F(r)}+
\frac{ r^2}{4} [(\sigma^1)^2+(\sigma^2)^2+s(r)^2(\sigma^3)^2] \ ,
\label{eq:MPBH}
\end{equation}
where $\sigma^a\,(a=1,2,3)$ are the invariant forms of $SU(2)$ defined as
\begin{align}
  \sigma^1 &= -\sin\psi d\theta + \cos\psi\sin\theta d\phi\ ,\notag \\
  \sigma^2 &= \cos\psi d\theta + \sin\psi\sin\theta d\phi\ , \notag \\
  \sigma^3 &= d\psi + \cos\theta d\phi  \ ,
\label{forms}
\end{align}
and the angular coordinate ranges are  $0\leq \theta < \pi $, $0\leq \phi <2\pi$, $0\leq \psi <4\pi$.
It is easy to show that the relation 
$d\sigma^a = \frac{1}{2} \epsilon^{abc} \sigma^b \wedge \sigma^c$ holds. 
In terms of these $SU(2)$ invariant forms, the metric of the round $S^3$ is
written as
\begin{equation}
 ds^2_{S^3}= \frac{1}{4} \{(\sigma^1)^2+(\sigma^2)^2+(\sigma^3)^2\}\ .
\end{equation}
On the other hand, in our metric ansatz~(\ref{eq:MPBH}), 
coefficients of $\sigma^1$, $\sigma^2$ and $\sigma^3$ are different if
$s(r)\neq 1$. Hence, the $t,r=\text{constant}$ surfaces (including the horizon) are
regarded as squashed $S^3$.

Now, we investigate the solution of the Einstein equation,
\begin{equation}
 R_{\mu\nu}-\frac{1}{2}g_{\mu\nu}R+6\lambda g_{\mu\nu}=0\ ,
\end{equation}
where $\lambda$ is defined as $\lambda=\Lambda/6$ and $\Lambda$ is a cosmological constant. 
The complete set of Einstein equations are
\begin{align}
&F'=
-\frac{2(
4\lambda r^3 ss'
+r^2F{s'}^2
-4rss'
+6\lambda r^2s^2
+3rFss'
+2rs^3s'
+5s^4
+3Fs^2
-8s^2)}
{rs(3s+rs')} \ ,
\label{eq:F}
\\
&s''=
\frac{
4\lambda r^3 ss'
+Fr^2{s'}^2
-4rss'
-rFss'
+2rs^3s'
+4s^4-4s^2}
{r^2Fs} \ ,
\label{eq:S}
\\
&\delta'=
-\frac{
4s^4
+2rs^3s'
-4s^2
+rFss'
-4rss'
+4\lambda r^3ss'
+r^2F{s'}^{2}}
{rFs(3s+rs')} \ ,
\label{eq:del}
\end{align}
where $'\equiv d/dr$.

\section{Numerical Solutions}
\label{Sec:Numerical}
We solve the ordinary differential equations~(\ref{eq:F})--(\ref{eq:del}) numerically in this section. 
We construct two types of solution: a warped black hole solution whose horizon and
AdS boundary are squashed $S^3$
and a warped AdS solution without a black hole.\footnote{
To the best of our knowledge, these solutions are different from the generalized Kerr-NUT-AdS 
solution.\cite{CLP, Houri1, Houri2}
We thank Yukinori Yasui for useful discussions on this issue.
} 

\subsection{Warped AdS Black Hole Solution}
\label{Sec:NumBH}
We construct an AdS black hole solution whose horizon is a squashed $S^3$ in this section.
The horizon is located at $F(r=r_+)=0$.
To solve the ordinary differential equations~(\ref{eq:F})--(\ref{eq:del}), 
we also have to supply boundary conditions to $s(r)$, $s'(r)$ and $\delta(r)$.
We can freely specify the boundary values $s(r=r_+)\equiv s_H$ and $\delta(r=\infty)$.
The latter can be set as an arbitrary value by redefining $t$, so we set it as $\delta(\infty)=0$.
$s'(r=r_+)$ is determined by imposing the regularity condition on the horizon.
That is, the right-hand side of (\ref{eq:S}) must be regular at $r=r_+$. 
Because the denominator of the right-hand side  
becomes zero at the horizon, the numerator must also be zero. 
Thus, the regularity condition is given by
\begin{equation}
 s'(r_+)=
\frac{2s_H(s_H^2-1)}{r_+(2-s_H^2-2\lambda r_+^2)}\ .
\label{sp}
\end{equation}
Thus, the solutions have two degrees of freedom: $(r_+, s_H)$, which are 
the horizon radius and the squashing parameter at the horizon, respectively.
These boundary conditions specify the series solution at $r=r_+$ to be
\begin{align}
F(r_++\epsilon)&=F'(r_+)\epsilon + \CO(\ep^2)\  ,\notag \\
s(r_++\epsilon)&=s_H + s'(r_+)\epsilon + \CO(\ep^2)\ ,\notag\\
s'(r_++\epsilon)&=s'(r_+)+s''(r_+)\epsilon + \CO(\ep^2)\ , 
\end{align}
where $F'(r_+)$ and $s''(r_+)$ are given by (\ref{eq:F}) and (\ref{eq:S}) as
\begin{gather}
s''(r_+)=
\frac{2s_H(s_H^2-1)(3s_H^4+14\lambda r_+^2s_H^2
-12s_H^2+8+8\lambda^2r_+^4-16\lambda r_+^2)}
{r_+^2(-2+s_H^2+2\lambda r_+^2)^3}\ , \\
 F'(r_+)=\frac{2(2-s_H^2-2\lambda r_+^2)}{r_+}\ .
\end{gather}

\begin{figure}[htbp]
 \centering
 \includegraphics[width=14cm, clip]{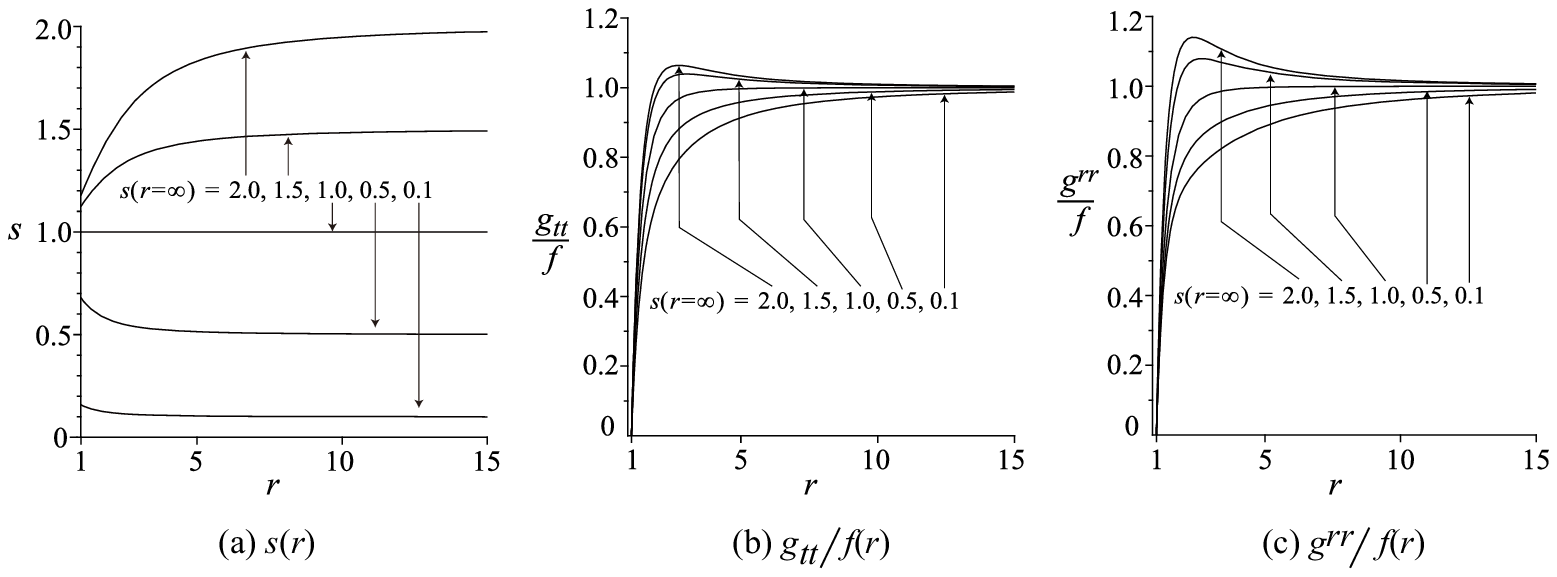}
 \vspace{-0.8cm}
 \caption{Warped AdS black hole solution for $r_+=1$ and
 $s(r=\infty)=2.0$, $1.5$, $1.0$, $0.5$ and $0.1$
in the unit of $\lambda=-1$. 
Panels (a)--(c) show the values of $s(r)$, 
$g_{tt}/f(r)$ and  $g^{rr}/f(r)$, respectively, where $f(r)=1-\lambda r^2$. The curve for $s(\infty)=1$ is 
the AdS-Schwarzschild black hole solution, which is not warped.
}
 \label{Fig:solutions}
\end{figure}

\begin{figure}[htbp]
 \centering
 \includegraphics[width=7cm, clip]{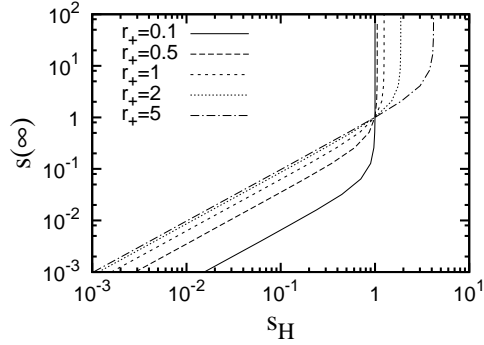}
 \caption{
 Relationship between $(r_+, s_H)$ and $s(\infty)$. For each value of $r_+$, 
 $s(\infty)$ monotonically increases from zero as we increase $s_H$ from zero, and it
 diverges for a critical value of $s_H$. A curvature singularity appears on such a point of 
 divergence. Thanks to this property, we can set $s(\infty)$ to an arbitrary value by tuning $s_H$.}
 \label{Fig:sH-sInf}
\end{figure}

Equations~(\ref{eq:F})--(\ref{eq:del}) are easily integrated numerically 
using these series solutions as initial conditions at $r=r_+$. 
In Fig.~\ref{Fig:solutions},
we show a solution for 
$r_+=1$ in the unit of $\lambda=-1$.
We plot the value of $s(r)$, as well as the ratios of $g_{tt}$, $g^{rr}$ to 
those of non-warped AdS spacetime, i.e.,
\begin{equation}
 g_{tt}/f(r)=F(r)e^{-2\delta(r)}/f(r)\ , \quad
g^{rr}/f(r)=F(r)/f(r)\ ,
\end{equation}
where $f(r) \equiv 1-\lambda r^2$. As we can see in Fig.~\ref{Fig:solutions},
these functions behave as
$s(r)\to\text{const.}$, $g_{tt}\to r^2$ and $g^{rr}\to r^2$ for 
$r\to\infty$. 
In general, $s(r)$ does not approach one but 
some other constant, and thus
the AdS boundary of the warped AdS black hole is a squashed $S^3$.
We also show in Fig.~\ref{Fig:sH-sInf} how $s(r=\infty)$ is determined using $(r_+, s_H)$.
We find that $s(\infty)$ appears to be a monotonically increasing function 
of $s_H$ for a fixed value of $r_+$, as shown in Fig.~\ref{Fig:sH-sInf}, 
and it diverges at some critical value of  $s_H$. When $s_H$ is larger than this critical value, 
$s(r)$ diverges at finite $r$.
We can show that a curvature singularity appears at such a point where $s(r)$ diverges.
Hence, the parameter range of $s_H$ for a regular solution is limited to the smaller value below 
a critical value, which is determined using $r_+$, and we can set $s(\infty)$ to an arbitrary value 
by tuning $s_H$. 
%

\subsection{Warped AdS Solution}
\label{Sec:solution}
\begin{figure}[htbp]
 \centering
 \includegraphics[width=14cm, clip]{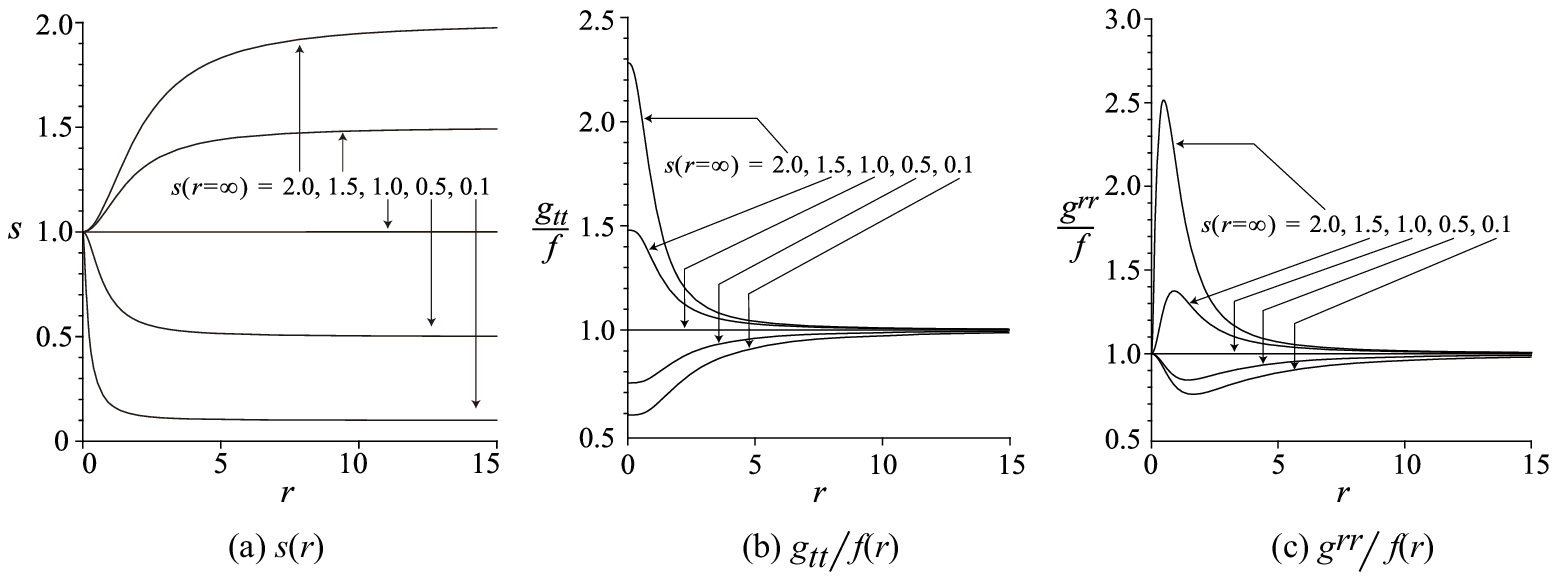}
 \caption{Warped AdS solutions for $s(\infty)=2.0$, 
 $1.5$, $1.0$, $0.5$ and $0.1$ in the unit of $\lambda=-1$.
 By tuning the free parameter $s_2$, we can set $s(\infty)$ to arbitrary values.
 This solution will be a background solution of a warped AdS black hole,
 which shares the same squashing parameter at the infinity 
 $s(\infty)$ with this solution.}
 \label{Fig:solutions_vac}
\end{figure}

\begin{figure}[htbp]
 \centering
 \includegraphics[width=7cm, clip]{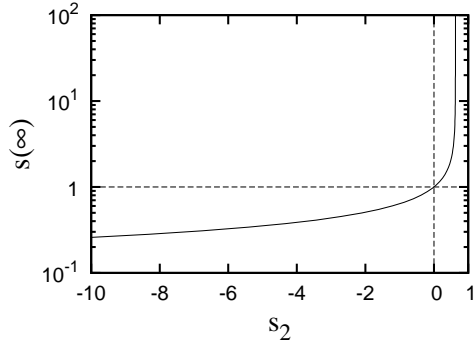}
 \caption{
 Relationship between $s_2$ at the origin and $s(\infty)$. 
 $s(\infty)$ monotonically increases as we increase $s_2$, and it
 diverges for a critical value of $s_2$. 
 Thus, we can set $s(\infty)$ to an arbitrary value by tuning $s_2$.
 $s(\infty)$ becomes one when $s_2=0$.
 For this parameter, the solution becomes a nonwarped AdS$_5$ 
 spacetime.}
 \label{Fig:s2-sInf}
\end{figure}

In this section, we construct a solution squashed at the AdS boundary without a black hole.
We call it  a warped AdS solution in this paper.
Such a solution can be constructed if we supply a boundary condition
at the origin instead of the horizon.
To avoid a conical singularity at the origin, we have to set 
$s(r=0)=1=F(r=0)$. The boundary value of $\d(r)$ can be set freely by 
rescaling $t$, so we set it to be $\d(r=\infty)=0$ as we did in the construction of the black hole solution.
The series solution at $r=0$, which satisfies these boundary conditions, is given by
\begin{align}
 s(0+\ep) &= 1 + \frac{s_2}{2} \ep^2 
    + \frac{s_2\left(4s_2+3\lambda\right)}{8}\ep^4+\mathcal{O}(\ep^5) \ , \\
 F(0+\ep) &= 1 - \left(s_2+\lambda\right)\ep^2 
    - \frac{s_2\left(9s_2+7\lambda\right)}{12}\ep^4 +\mathcal{O}(\ep^5)\ .
\end{align}
$s_2$ in this expression can be set freely, so this solution has one
degree of freedom.
We show some solutions in Fig.~\ref{Fig:solutions_vac}.\footnote{
We can introduce magnetic charge $M$ by coordinate
transformation $\psi=-x^5/M-\phi$ and redefinition of the variable
$s(r)=M\tilde{s}(r)$. In the limit of $M\to 0$, 
the warped AdS solution does not reduce to the nonwarped AdS
spacetime,\cite{Onem} but if we take another limit $s_2\to 0$, 
this solution reduces to nonwarped AdS solution.
}
We can see that $s(r)$, $g_{tt}$ and $g^{rr}$ behave as 
$s(r)\to\text{const}$, $g_{tt}\to r^2$ and $g^{rr}\to r^2$ for 
$r\to\infty$ in these solutions,
like they do in the warped AdS black hole solutions.
We also show the relationship between $s_2$ at the origin and $s(\infty)$ in 
Fig.~\ref{Fig:s2-sInf}.
$s(\infty)$ diverges at a critical value of $s_2$,
as it did in the warped AdS black hole case for a critical value of $s_H$ 
(see Fig.~\ref{Fig:sH-sInf}), so 
we can set $s(\infty)$ arbitrarily by tuning $s_2$.
This warped AdS solution will be a vacuum or background solution for 
the warped AdS black hole solution which has the same squashing
parameter at the infinity $s(\infty)$, similarly to the nonwarped pure AdS spacetime
for AdS-Schwarzschild black holes.

\subsection{Temperature and Entropy}
\label{Sec:S-T}
In this section, we summarize the expressions of temperature and entropy 
of the warped AdS black hole solution we constructed in \S\ref{Sec:NumBH}.
We also illustrate the behavior of the temperature and the entropy of the warped AdS black holes and observe 
their similarity to those of AdS-Schwarzschild black holes.

The temperature of the warped AdS black hole is given by
\begin{equation}
 T\equiv\frac{\kappa}{2\pi}=
\frac{F'(r_+)e^{-\delta(r_+)}}{4\pi}
=\frac{(2-s_H^2-2\lambda r_+^2)e^{-\delta(r_+)}}{2\pi r_+}
\ ,
\end{equation}
where $\kappa$ is surface gravity of $\partial_t$. 
We note here that the definition of temperature has an ambiguity that comes from the 
arbitrariness of the scale of time-coordinate $t$; we can freely rescale $t$ as $t'\equiv Ct$
and if we do that
the temperature will be rescaled as $T'= T/C$. 
We fix this ambiguity by defining the temperature using always the metric~(\ref{eq:MPBH})
with $\delta(\infty)=0$.
This convention facilitates the comparison of the gravity theory to the gauge theory on the warped AdS boundary,
whose conformal metric is given by (\ref{Sq}).

The entropy of the warped AdS black hole is given by
\begin{equation}
 S\equiv\frac{\text{Area(horizon)}}{4 G_5}=\frac{\pi^2r_+^3s_H}{2G_5}\ ,
\end{equation}
where $G_5$ is a five-dimensional Newton constant and can be written as 
$1/G_5=2N^2/(\pi \ell^3)$ where $\ell=(-\lambda)^{-1/2}$ and 
$N$ is the number of colors in the dual $U(N)$ gauge
theory. 

\begin{figure}[htbp]
    \centering
    \includegraphics[width=7cm, clip]{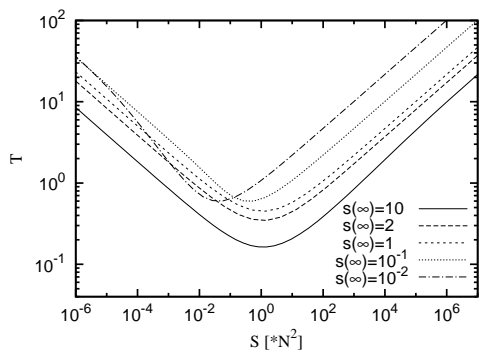}
    \caption{
    Relationship between $S/N^2$ and $T$ of warped AdS black holes for 
    $s(\infty)=10$, $2$, $1$, $10^{-1}$ and $10^{-2}$ in the unit of $\lambda=-1$.
    $T$ has a minimum for any $s(\infty)$, and its position is determined using the value of $s(\infty)$.
    }
    \label{Fig:S-T_grav}
\end{figure}

We show the relationship between $T$ and $S$ of the warped AdS black holes 
for various $s(\infty)$ in Fig.~\ref{Fig:S-T_grav}. 
For any $s(\infty)$, $T$ has a minimum at some value of $S$ as in 
the case of an AdS-Schwarzschild black hole.

Since the properties of $T$ and $S$ are quite similar to those of an AdS-Schwarzschild black hole, we 
expect that the Hawking-Page transition~\cite{HP} takes place even for our warped AdS black hole, 
and the transition temperature is approximated using the temperature that minimizes the entropy.
However, we could not show that it is really the case because it is not straightforward to calculate
numerically the mass and the free energy of the warped black holes.\footnote{
The properties of the mass and the free energy are investigated in Ref.~\citen{Bri}.
}

\section{Dual Gauge Theory}
\label{Sec:Gauge}
In this section, we study the gauge theory that lives on the 
spatial infinity of the warped AdS black hole. 
It is expected to be $U(N)$ $\CN=4$ super Yang-Mills theory
on a squashed $S^3$, because the warped solution becomes the ordinary
AdS space when the squashing parameter is set to one, where we have 
the well-known AdS/CFT correspondence.

First, we review the spectrum in this theory in \S\ref{Sec:spectrum}.
Then we calculate the effective action and 
study the phase structure of this theory in \S\ref{Sec:EA}.
The confinement/deconfinement transition occurs for any squashing parameter 
as long as the theory is well defined.
We clarify the dependence of the transition temperature on the squashing parameter and show 
that the transition temperature becomes zero if the theory no longer applies.
After that, we calculate the entropy of the gauge theory in \S\ref{Sec:compare}
 and compare it with that of the gravity theory.
We find a good agreement between the entropies of the two theories and obtain the well-known 
ratio $3/4$ for any squashing parameter $s$.

\subsection{Spectrum}
\label{Sec:spectrum}
Consider $\CN =4$ SYM on a squashed $S^3$.
The action is
\begin{align}
	I_\text{gauge} = -\int d^4 x \s{-g}\, \text{tr} \Big[ \f{1}{2}(F_{\m\n})^2 + (D_\m\phi_m)^2
	+& \f{R}{6}\phi_m^2 + i\bar\lambda^A\G^\m D_\m\lambda_A \no
	 &- \f{g^2}{2}[\phi_m, \phi_n]^2 - g\bar\lambda^A\G^m [\phi_m, \lambda_A 		] \Big] \ ,
\label{SYMaction}
\end{align}
where $\m= 0,1,2,3,\ m=1,2,\dots,6,\ A=1,\dots,4$, 
$F_{\mu\nu}=\partial_\mu A_\nu-\partial_\nu A_\mu+ig[A_\mu,A_\nu]$ and
$A_\mu$ is the gauge field of $U(N)$. 
$\phi_m$ is a scalar field and $\lambda_A$,
which is originally a gaugino in the $\bold{\Bar{16}}$ representation 
of ten-dimensional type IIB 
supergravity,
is a four-dimensional spinor in the $(\bold{2}, \bold{\bar 4}) 
+ (\bold{\bar 2},\bold{4})$
representation under $SO(1,3)\times SU(4)$.
All fields are adjoint representation of $U(N)$. 
The gauge covariant derivative is defined as
$D_\mu = \nabla_\mu + ig[A_\mu,\ \cdot\ ]$.
$R$ in the mass term of $\phi_m$ is the Ricci scalar of the background
geometry. 
For simplicity, we will consider the zero coupling limit of this theory 
except for the zero mode of the
$A_0$ as in Ref.~\citen{AMMPV}.

In the gravity theory, 
the geometry of $r=r_0$ surface at the AdS boundary
of the warped AdS black hole solution
is given by 
\begin{equation}
 ds^2\simeq \frac{r_0^2}{\ell^2}\left[
-dt^2+\frac{\ell^2}{4}\{ (\sigma^1)^2 + (\sigma^2)^2 + s^2(\sigma^3)^2 \}
\right] \quad \text{for}\quad r_0\to \infty\ ,
\label{rinf}
\end{equation}
where $\ell\equiv \sqrt{-6/\Lambda}$ and $s\equiv s(r=\infty)$.
We consider the $\CN =4$ SYM theory on this background~(\ref{rinf}).
The conformal factor $r_0^2/\ell^2$ can be neglected 
because of the conformal symmetry of the theory, and
thus the background metric can be reduced to 
\begin{align}\label{Sq}
  ds^2 = -dt^2 + \f{\ell^2}{4}\{ (\sigma^1)^2 + (\sigma^2)^2 + s^2(\sigma^3)^2 \} \ .
\end{align}
We take the unit of $\ell=1$. 
In this unit, the Ricci scalar is given as $R=2(4-s^2)$ in terms of $s$.
On the background~(\ref{Sq}), the gauge theory~(\ref{SYMaction}) has no
 supersymmetry as shown in the Appendix.
The dual vectors of $\sigma^a$, defined as $\sigma^a_i\,e^i_b = \d^a_b$,
are given by
\begin{align}
  e_1 &= -\sin\psi\,\p_\t + \f{\cos\psi}{\sin\t}\,\p_\phi -
  \cot\t\,\cos\psi\, \p_\psi \ ,\no
  e_2 &= \cos\psi\,\p_\t + \f{\sin\psi}{\sin\t}\,\p_\phi -
  \cot\t\,\sin\psi\,\p_\psi \ ,\no
  e_3 &= \p_\psi \ . 
\end{align}
In addition, there are Killing vectors that characterize the $SU(2)$
symmetry of the metric (\ref{Sq}):
\begin{align}
  \xi_x &= \cos\phi\,\p_\t + \f{\sin\phi}{\sin\t}\,\p_\psi -
  \cot\t\,\cos\phi\, \p_\phi \ ,\no
  \xi_y &= -\sin\phi\,\p_\t + \f{\cos\phi}{\sin\t}\,\p_\psi -
  \cot\t\,\cos\phi\,\p_\phi \ ,\no
  \xi_z &= \p_\phi \ . 
\end{align}

Let us consider the spectrum of the fields on a squashed $S^3$
(\ref{Sq}).
We can find that the metric has $SU(2)\times U(1)$ symmetries
that are generated by the Killing vectors $\xi$'s and $e_3$, respectively.
Then, all fields on the squashed $S^3$ can be decomposed into
the irreducible representation of $SU(2)\times U(1)$.
Let us introduce two types of angular momentum operators 
\begin{align}
  L_\a = i\xi_\a \ ,\qquad W_a = ie_a \ .
\end{align}
These satisfy commutation relations 
\begin{align}\label{Com}
  [L_\a, L_\b] = i\ep_{\a\b\g}L_\g \ ,\qquad [W_a, W_b] =
  -i\ep_{abc}W_c \ , \qquad [L_a,W_a]=0 \ . 
\end{align}
The commutation relations (\ref{Com}) show that we can
simultaneously diagonalize $L^2,\,L_z$ and $W_3$, then all the states 
can be completely specified using eigenvalues of these operators.
For scalar fields, the eigenstate
 is defined as
\begin{align}
  L^2|J,M;K\rangle &= J(J+1)|J,M;K\rangle \ ,\no
  L_z|J,M;K\rangle &= M|J,M;K\rangle \ ,\no
  W_3|J,M;K\rangle &= K|J,M;K\rangle \ ,
\end{align}
%
where $J$ takes all the values of half-integer, and $M$ and $K$ take the
value of
$K,M=-J,-J+1,\cdots,J$. 
The energy of a scalar field that conformally couples to a squashed
$S^3$ is obtained as~\cite{Hu}
\begin{align}
  E_S = \s{(2J+1)^2 + \a \left( 4K^2 + \f{1}{3(1+\a )} \right)} \ ,
\label{Es}
\end{align}
where $\a \equiv 1/s^2 - 1$.
This expression reproduces the energy of a scalar field on the round $S^3$
in the limit of $s\to 1$.
We must mention that  
the terms in the square root of this energy spectrum become negative when $s>2$ for some $J, K$: 
$J=0=K$ for example.
In this case, the energy spectrum becomes tachyonic and the theory cannot be defined properly. 
The appearance of tachyon is due to the conformal scalar mass term $R\phi^2_m/6=3(4-s^2)\phi_m^2$
in the action~(\ref{SYMaction}), 
which becomes negative for $s>2$.
Hence, we have to set the squashing parameter $s$ 
to be in the region $0<s\le 2$ so that the theory is well defined.

For spinors, we must be careful about their chirality.
A spinor with a positive chirality can be represented as
$|J,M;K\rangle$ with $M= -J,-J+1,\cdots,J$ and 
$K=-J-1/2,-J+1/2,\cdots,J+1/2$.
Its energy is obtained as~\cite{Gi} 
\begin{align}
  E_F^{(+)} = \f{s}{2} + 2\s{\left(J+\f{1}{2}\right)^2 + \a K^2} \ .
\end{align}
Similarly, a spinor with a negative chirality can be represented as
$|J,M;K\rangle$ with 
$M=-J-1/2,-J+1/2,\cdots,J+1/2$
and
$K= -J,-J+1,\cdots,J$,
and 
its energy is
\begin{align}
  E_F^{(-)} = -\f{s}{2} + 2\s{(J+1)^2 + \a K^2} \ .
\end{align}
We note here that this $E_F^{(-)}$ becomes negative only for $s>4$, where small-$J$ 
modes become negative first for such a large $s$.
 
The spectrum of vector fields is also obtained in Ref.~\citen{Gi}, 
and it is given by
\begin{align}\label{Ev}
  E_V^{(+)} = s + 2\s{\left(J+\f{1}{2}\right)^2 + \a \left( K^2 -
      \f{l^2}{4} \right)} \ , 
\end{align}
for positive helicity representation $|J,M;K\rangle$ 
with $M= -J,-J+1,\cdots,J$ and
$K= -J-1,-J,\cdots,J+1$, and 
\begin{align}
  E_V^{(-)} = -s + 2\s{\left(J+\f{3}{2}\right)^2 + \a \left( K^2 -
      \f{l^2}{4} \right)} \ , 
\end{align}
for negative helicity representation
$|J,M;K\rangle$ with $M= -J-1,-J,\cdots,J+1$
and $K= -J,-J+1,\cdots,J$.
This $E_V^{(-)}$ is always positive for any $s$ and any mode.

\subsection{Effective Action and Phase Transition}
\label{Sec:EA}
We summarize how to calculate the effective action of this gauge theory on a squashed $S^3$
and then show the phase structure of the theory.
Especially, the confinement/deconfinement transition occurs in this case.
We illustrate the dependence of the transition temperature on the squashing parameter.

The expression of the effective action of free $\CN=4$ $U(N)$ SYM theory is 
originally given in Refs.~\citen{Su} and \citen{AMMPV}
utilizing a unitary matrix model (see also Refs.~\citen{YY} and \citen{MNTY}).
The derivation is parallel even for a squashed $S^3$ background, so we skip it.
The difference from the round $S^3$ background case only appears in the energy spectrum, 
which was studied in \S\ref{Sec:spectrum}.

The effective action $I_\text{gauge}$ of the free $\CN=4$ $U(N)$ SYM theory on a squashed $S^3$
is given by
\begin{equation}
 I_\text{gauge}[\rh(\t)] = N^2 \sum^\infty_{n=1} \rh_n^2 V_n \ ,
  \label{Igauge}
\end{equation}
where $\rh_n\equiv \int^\pi_{-\pi}d\t \rh(\t)\cos (n\t)$ and 
$\rh(\t)$ is the eigenvalue density distribution of $U(N)$ generators in
the large-$N$ limit.
Namely, $\rho(\t)$ is the eigenvalue density of the Polyakov loop around
the thermal cycle, and the effective action (\ref{Igauge}) can be
obtained after integrating out all harmonic modes on a squashed $S^3$
from the $\CN =4$ SYM action (\ref{SYMaction}).
$V_n$ in the expression is given by 
\begin{equation}
 V_n \equiv \frac{1}{n}\left(1-(z_S(x^n)+z_V(x^n))-(-1)^{n+1}z_F(x^n)\right)\ ,
\end{equation}
where $x\equiv e^{-1/T}$. 
$z_S$, $z_F$ and $z_V$ are the single-particle partition functions of scalar, 
spinor and vector field, respectively, whose explicit expressions are
\begin{align}
 z_S(x) &= 6\sum_{J=0,1/2,\dots}^\infty \sum_{M=-J}^J \sum_{K=-J}^J x^{E_S}\ , \notag \\
 z_F(x) &= 4\sum_{J=0,1/2,\dots }^\infty \sum_{M=-J}^J \sum_{K=-J-1/2}^{J+1/2} x^{E_F^{(+)}}
 + 4\sum_{J=0,1/2,\dots }^\infty \sum_{M=-J-1/2}^{J+1/2} \sum_{K=-J}^{J}  x^{E_F^{(-)}}\ ,
 \no
 z_V(x) &= \phantom{1}
 \sum_{J=0,1/2,\dots }^\infty \sum_{M=-J}^J \sum_{K=-J-1}^{J+1} x^{E_V^{(+)}}
 \;\; + \phantom{1}
 \sum_{J=0,1/2,\dots }^\infty \sum_{M=-J-1}^{J+1} \sum_{K=-J}^{J} x^{E_V^{(-)}}\ .
\end{align}
The expression of the effective action~(\ref{Igauge}) shows that the effective 
action is minimized when $\rh_n=0$ for all $n$, which corresponds to 
the uniform distribution of $\rh(\t)$,  if the inequality
\begin{equation}\label{uniform}
 V_n>0\quad \Leftrightarrow \quad
z_S(x^n)+z_V(x^n)+(-1)^{n+1}z_F(x^n)<1 \qquad \text{for all }n
\end{equation}
is satisfied. 
Since the single-particle partition functions are monotonically 
increasing functions of $x$ and $x$ takes its value in the range $0<x<1$, the
inequality for $n=1$ is firstly violated as we increase $x$ from zero, 
i.e., as we increase the temperature from zero.
Hence, above the critical temperature $T_H$ marked by
\begin{equation}
 z_S(x_H)+z_V(x_H)+z_F(x_H)=1\ ,
\end{equation}
where $x_H=e^{-1/T_H}$,
the uniform distribution of $\rh(\t)$ does not minimize the effective action.
It can be shown that the effective action becomes $\CO(N^2)$ above this 
temperature, while below which it is $\CO(1)$. This transition is known as a 
confinement/deconfinement transition of gauge theory.

\begin{figure}[t]
 \centering
 \includegraphics[width=7cm, clip]{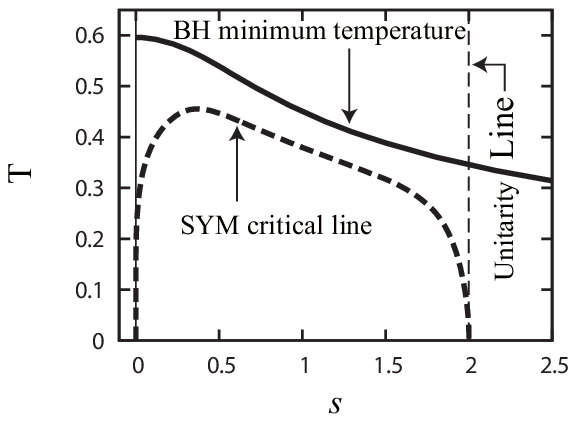}
 \caption{Phase diagram of the $\CN=4$ $U(N)$ SYM theory on a squashed $S^3$ 
 whose squashing parameter is $s$. The thick dashed line shows the critical 
 temperature $T_H$ of the SYM theory.
 The phases below and above this critical line 
 are the confinement and deconfinement phases, respectively.
 The critical temperature $T_H$ becomes zero in the limit of $s\to 0$ and 
 $s\to 2$, and the SYM theory becomes tachyonic and its unitarity is violated for $s>2$.
 The thick solid line shows the minimum temperature of the warped AdS black holes for 
 fixed values of $s=s(\infty)$; this line is expected to be close to the Hawking-Page 
 transition line of the warped AdS spacetime. The behavior of this line is 
 qualitatively similar to that of the SYM critical line, at least for $s\sim 1$.
}
 \label{Fig:phase}
\end{figure}

In Fig.~\ref{Fig:phase}, we draw the phase diagram of the gauge theory 
on a squashed $S^3$ for each squashing parameter $s$. The phases below and above 
the critical line correspond to the confinement and deconfinement phases, respectively. 
When $s=1$, we reproduce the result of Ref.~\citen{AMMPV}
for a round $S^3$: $T_H\sim 0.3797\,$. 
We see from the figure that the critical temperature $T_H$ is close to this value when
$s$ is close to one, and it goes to zero in the limit $s\to 0$ and $s\to 2$.
We can read off it analytically for $s=0$ and $s=2$ from the spectra (\ref{Es}) and (\ref{Ev}),
respectively, where the energy of the lowest mode becomes zero, and the condition for confinement 
(\ref{uniform}) is no longer satisfied even at zero temperature.
Therefore, this transition takes place for any $s$ as long as the theory
is well defined 
(i.e., $0<s\le 2$), 
and it goes to zero in the limit the theory breaks down.
We discuss the implications of this phase structure to the gravity theory in \S\ref{Sec:discussion}.

In Fig.~\ref{Fig:phase}, we also plot the minimum temperature of the 
warped AdS black holes for each squashing parameter $s(\infty)$.
As we mentioned in \S\ref{Sec:S-T}, this temperature is expected to be 
close to the Hawking-Page transition temperature of warped AdS black holes.
From Fig.~\ref{Fig:phase}, we can see a qualitative similarity of this 
line and the SYM critical line, at least for $s\sim 1$, while they behave differently 
for $s>2$. We  discuss the implications of these similarities and dissimilarities 
in \S\ref{Sec:discussion}.

\subsection{Entropy and Comparison with Dual Gravity}
\label{Sec:compare}
In this section, we calculate the entropy of the gauge theory and then compare 
it with that of the warped AdS black hole solution we 
constructed in \S\ref{Sec:solution}.

In Refs.~\citen{Su} and \citen{AMMPV}, the exact solution of the effective action $I_\text{gauge}$ 
is obtained for $T>T_H$ in the large-$N$ limit, 
and it can be approximated as follows if 
$z_n(x) \equiv z_S(x^n) + z_V(x^n) + (-)^{n+1}z_F(x^n)$ 
decreases exponentially with $n$ for $n>1$:
\begin{gather}\label{ApDenFunc}
  \rho(\t) = 
  \begin{cases}
    \s{\sin^2\left(\f{\t_0}{2}\right) - \sin^2\left(\f\t
        2\right)} \cos\f{\t}{2} 
    \> \big/ \>
    \pi\sin^2\left(\f{\t_0}{2}\right)
    & \left(\left|\t\right|<\t_0\right)
    \\ 
    \qquad \qquad \qquad \qquad \;\;  0 
    & \left( \text{elsewhere} \right)
  \end{cases} \quad,
  \\
  \sin^2\left(\f{\t_0}{2}\right) = 1- \s{1-\f{1}{z_1(x)}}\quad.
\end{gather}
The factor $z_n$ does in fact decrease exponentially, 
and thus we can use (\ref{ApDenFunc}) as a good approximation.
Substituting (\ref{ApDenFunc}) into (\ref{Igauge}), we obtain the
effective action in a very simple form:
\begin{align}\label{actCFT}
  I_{\text{gauge}}= -N^2\left( \f{1}{2\sin^2 \left( \f{\t_0}{2} \right)} +
    \f{1}{2}\log \left(  \sin^2\left( \f{\t_0}{2} 
      \right) \right) -\f{1}{2} \right) \ .
\end{align}
The entropy is obtained from this effective action as 
\begin{equation}
 S_\text{gauge} = \left(x\log x\frac{\p}{\p x}-1\right)I_\text{gauge}\ .
\end{equation}

\begin{figure}[htbp]
 \centering
 \includegraphics[width=7cm, clip]{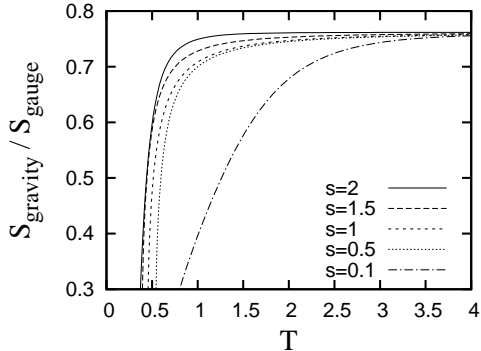}
 \caption{Ratio of the entropy of SYM to that of the dual gravity.
 The ratio becomes about $3/4$ irrespective of the squashing parameter $s$
 even at low temperature.
 For smaller $s$, the convergence of the ratio into $3/4$ is 
achieved at higher temperature. 
 }
 \label{Fig:compare}
\end{figure}

In Fig.~\ref{Fig:compare}, we show the ratio of the entropies of the gauge 
theory and the dual gravity theory.
They match well irrespective of the squashing parameter $s$, and the ratio 
takes a well-known value $3/4$,
even in the limit at which the gauge theory no longer applies ($s \to 2$).
This ratio is expected to become $1$ if we 
take the strong-coupling effect into account.
In the low-temperature region, the ratio deviates from $3/4$ significantly, but 
it merely comes from the difference of the transition temperature
between the two theories.
Thus we expect that a duality exists between the warped AdS spacetime and $\CN =4$ SYM 
theory on the squashed $S^3$ irrespective of $s$ as long as the theory
is well defined.

We must mention that the more we decrease the squashing parameter $s$ to zero, 
the more we have to increase the temperature to achieve the convergence 
of the entropy ratio to $3/4$. 
It may imply that the coincidence between the 
entropies of the two theories will be lost in the limit $s\to 0$, 
i.e., the duality breaks down in this limit.
On the other hand, the ratio is always approximately 3/4 at high temperature for any $s$. 
This is a physically consistent result because the squashing can be neglected 
and the duality goes to the ordinary AdS/CFT correspondence in the
Poincar{\' e} coordinate
at the high-temperature region.

\section{Discussion}
\label{Sec:discussion}

In this paper, we constructed numerical solutions of warped AdS black holes 
as well as warped AdS spacetime without black holes
and considered $\CN =4$ super Yang-Mills theory on a squashed $S^3$, which is expected 
to be dual to the gravity theory in warped AdS spacetime.
This duality between the gravity on warped AdS spacetime and the SYM on a 
squashed $S^3$ can be regarded as an extension of the usual AdS/CFT duality between 
the gravity on nonwarped AdS and the SYM on a round $S^3$.
We found some circumstantial evidence of this duality,
although there is no supersymmetry in the gauge 
theory as shown in the appendix.

The entropies of the two theories matched up to a factor of $3/4$ 
irrespective of the squashing parameter $s$, as long as $0<s\le 2$ and
the gauge theory is well defined.
This fact suggests that the warped AdS/CFT correspondence works in this parameter region $0<s \le 2$.
However, the temperature at which the ratio converges into $3/4$ increases as we 
decrease $s$ to zero. It may indicate that the correspondence breaks down in this 
limit $s\to 0$.

As we showed in \S\ref{Sec:EA}, 
the gauge theory exhibits a  transition
between the confinement and deconfinement phases
as long as the theory is well defined.
In the gravity theory, we have a minimum temperature line that resembles this transition line at 
least for $s\sim 1$
as shown in Fig.\ref{Fig:phase}.
We expect that this minimum temperature line is very close to the Hawking-Page transition line
of the warped AdS black hole, since the transition temperature of a
nonwarped AdS-Schwarzschild black hole is 
close to its minimum temperature.
If this expectation is correct, 
this result indicates that the transition lines in the two theories resemble
each other qualitatively, and
this agreement provides  another evidence of the warped AdS/CFT correspondence.

Despite of these agreements, 
there are some discrepancies between these theories.
For $s>2$,
the gauge theory is poorly defined
while the warped AdS black hole does exist.
The minimum temperature line of the gravity theory in this parameter region $s>2$
does not resemble the critical line of the gauge theory.
This result suggests 
an interesting scenario such that 
the warped AdS black hole becomes unstable when 
the dual gauge theory becomes unstable, i.e., when $s>2$,
and settles down into some other new phase.
It will be interesting to investigate the dynamical and
thermodynamical stabilities of the warped
AdS black hole in this parameter region\footnote{
For some black holes that have the
same isometry as the warped AdS black hole, 
the stability was studied in Refs.~\citen{MS,KMSI,IKKMSSZ,MS2,Mura}.
} and determine
whether this duality between the warped AdS and the SYM on a squashed $S^3$ works 
in this parameter region $s>2$.

\section*{Acknowledgements}
We are grateful to H.~Kawai, T.~Kobayashi, H.~Ishihara, Y.~Yasui, D.~Ida, S.~Tomizawa,
T.~Tanaka, M.~Kimura
and Y.~Sumitomo for valuable discussions, and J.~Soda for collaboration at an earlier stage.
The works of KM, TN and NT are supported by JSPS Grants-in-Aid for Scientific Research
Nos.\,19$\cdot$3715, 19$\cdot$3589 and 20$\cdot$56381, respectively.
This work was supported by a Grant-in-Aid for the Global COE Program ``The Next Generation 
of Physics, Spun from Universality and Emergence'' from the Ministry of Education, Culture, 
Sports, Science and Technology (MEXT) of Japan.

\appendix
\section{Supersymmetry of $\CN =4$ SYM on Three-Dimensional Manifold}
We study $\CN = 4$ SYM on $R \times M_3$, where $M_3$ is an arbitrary
three-dimensional manifold with constant Ricci curvature.
The background metric is
\begin{align}
  ds^2 = -dt^2  + ds^2(M_3).
\end{align}
The action on this background 
\begin{align} 
	I = -\int d^4 x \s{-g}\, \text{tr} \Big[ \f{1}{2}(F_{\m\n})^2 + (D_\m\phi_m)^2
	+& \f{R}{6}\phi_m^2 + i\bar\lambda^A\G^\m D_\m\lambda_A \no
	 &- \f{g^2}{2}[\phi_m, \phi_n]^2 - g\bar\lambda^A\G^m [\phi_m, \lambda_A 		] \Big] \ ,
\end{align}
is invariant under the supersymmetric transformation
\begin{align}
	\d A_\m &= -i \bar\lambda^A \G_\m \ep_A \ , \qquad 
	\d \phi_m = -i \bar\lambda^A \G_m \ep_A \ ,  \no
	\d \lambda_A &= \left[ \f{1}{2}F_{\m\n}\G^{\m\n} - D_\m \phi_m\G^m \G^\m 	-\f{1}{2}\phi_m\G^m\G^\m\nabla_\m - \f{i}{2}[\phi_m, \phi_n ]\G^{mn}  \right]\ep_A \ ,
\end{align}
only when the spinors $\ep_A$ satisfy the Killing spinor equations
\begin{align}\label{Killing}
	& \left(  \p_0 - \f{1}{2}\s\f{R}{6}\g_0 \right)\ep = 0 \ , \qquad \left( \nabla_i - \f{1}{2}\s\f{R}{6}	
	\g_i\g_5 \right)\ep = 0 \ , \nonumber \\
	& \quad \nabla_i \ep \equiv \left( \p_i + \f{1}{4}\om^{ab}_i \g_{ab} \right)\ep \ .
\end{align}
They must also satisfy the integrability condition
\begin{align}\label{Integ}
	[\nabla_\m, \nabla_\n] \ep = \f{1}{4}R_{ij}^{~~ kl}\g_{kl}\ep \ ,
\end{align}
then we obtain the condition for preserving the supersymmetry putting
(\ref{Killing}) into (\ref{Integ}) as
\begin{align}
	R_{ij,kl}\g^{[kl]} = \f{R}{3}\g_{[ij]} \ .
\end{align}
Multiplying $\g^{mn}$ and taking trace on both sides, we obtain
\begin{align}
  R_{ij,kl} = \f{R}{6}(g_{ik}g_{jl} - g_{il}g_{jk}) \ ,
\end{align}
which indicates that $M_3$ must be a maximally symmetric space such 
as $S^3$, $R^3$ and $H_3$.
Thus $\CN =4$ SYM on the squashed $S^3$ does not have any supersymmetry.

\end{document}